# Hydrostatic pressure effects on the electrical transport properties of $Pr_{0.5}Sr_{0.5}MnO_3$

F. J. Rueckert,[1*] M. Steiger,[1] B. K. Davis,[1,2] T. Huynh,[3] J. J. Neumeier,[3] and M. S. Torikachvili[1]

[1] *Department of Physics, San Diego State University, San Diego, CA 92182*

[2] *Quantum Design, 6325 Lusk Boulevard, San Diego, CA 92121*

[3] *Department of Physics, Montana State University, Bozeman, MT 59717*

**Abstract**

We studied single-crystalline $Pr_{0.5}Sr_{0.5}MnO_3$ by means of measurements of magnetic susceptibility and specific heat at ambient pressure ($P$), and electrical resistivity ($\rho$) in hydrostatic pressures up to 2 GPa. This material displays ferromagnetic (FM) order, with Curie temperature $T_C \approx 255$ K. A crystallographic transformation from I4/mcm to Fmmm is accompanied by the onset of antiferromagnetism (AFM), with Néel temperature $T_N \approx 161$ K. The effect of pressure is to lower $T_C$, and raise $T_N$ at the approximate rates of -3.2 K/GPa, and 14.2 K/GPa, respectively. Although the value of $T_N$ increases with $P$, due to the enhancement of the superexchange interactions, the AFM-Fmmm state is progressively suppressed, as pressure stabilizes the FM-I4/mcm phase to lower temperatures. The $\rho$ vs $T$ data suggest that the AFM phase should be completely suppressed near 2.4 GPa.

**Introduction**

   The interplay between the magnetic and crystallographic structures in the perovskite manganites of general composition $R_{1-x}A_x$MnO$_3$, where $R$ is a trivalent rare earth ion and $A$ is a divalent alkaline element, results in a number of interesting features, including remarkable magnetic and electronic behavior, e.g. colossal magnetoresistance (CMR)[1,2] and charge order.[3] In light of the ability to tune these properties through experimental parameters, as for example composition, temperature ($T$), magnetic field ($H$), etc., it is tempting to consider these materials for a number of technological applications.[4-6] A charge ordered insulating state is frequently observed below a characteristic temperature $T_{co}$ in the manganites at half-doping level, i.e., when divalent $A$ substitutes half of trivalent $R$.[6,7] The regular arrangement of Mn$^{3+}$ and Mn$^{4+}$ ions, first described by Goodenough as charge ordering,[3] sets in at $T_{co}$, as long range Coulomb interactions overcome the kinetic energy of charge carriers, inhibiting the double-exchange mechanism.[8]

   Charge order and CMR are electronic phenomena originating from the interaction of charge carriers with both Mn$^{3+}$ and Mn$^{4+}$ ions and lattice distortions. Electronic and magnetic properties can be further affected by hole doping.[9] Electrical conductivity in these materials is sustained by a double-exchange mechanism as proposed by Zener,[10] in which charge flows as electrons hop between neighboring Mn ions with parallel spins along the Mn$^{3+}$-O-Mn$^{4+}$ bonds,[11] therefore providing coupling between electrical conductivity and spin dynamics.

   The Pr$_{0.5}$Sr$_{0.5}$MnO$_3$ compound shows remarkable electronic and magnetic properties. As it is cooled below the Curie temperature ($T_C \approx 270$ K in Ref. 6), the electrical resistivity ($\rho$) shows a noticeable drop due to the suppression of spin-flip scattering.[6] Upon further cooling, a sharp upward discontinuity in the $\rho$ vs $T$ data occurs near 140 K, below which the resistivity continues to increase. This discontinuity is reminiscent of a metal-insulator transition. Tomioka et al. first suggested that this transition is due to the onset of charge ordering. The application of an external magnetic field drove $T_{co}$ down, eventually "melting" the charge-ordered state completely for $\mu_0 H \approx 7$ T, and reestablishing the metallic state.[6] However, a neutron diffraction study of Pr$_{0.5}$Sr$_{0.5}$MnO$_3$ showed that there was no clear indication of charge ordering,[12] and that the phase transition at 140 K is due to the formation of an A-type antiferromagnetic (AFM) structure. Therefore, we will refer to the magnetic phase transition from FM to AFM as occurring at the Néel temperature ($T_N$). Similarly, neutron diffraction results by Damay et al. show no evidence of charge ordering near $T_N$.[13] They also show that the crystal structure of the ferromagnetic (FM) phase (tetragonal; space group I4/mcm) changes to orthorhombic (space group Fmmm) near $T_N$. The onset of this A-type AFM structure correlates well in temperature with the feature in their $\rho$ vs $T$ data near 140 K, consistent with the observation of a metal-insulator transition at this temperature.

It should be pointed out that metallic behavior in $\rho$ vs $T$ above $T_N \approx 140$ K was observed in single-crystals reliably[6] whereas data on polycrystalline $Pr_{0.5}Sr_{0.5}MnO_3$ showed approximately exponential activation above $T_N$.[14] Chen et al. observed a large change in entropy near 161 K in polycrystalline $Pr_{0.5}Sr_{0.5}MnO_3$ by means of measurement of magnetization *(M)* as a function of temperature.[7] The change in entropy of 7.1 J/kg-K near $T_N$ is positive and much greater than values measured near $T_C$. Chen et al. suggest that this large change in entropy at 161 K is due to a magnetic phase transition from FM to AFM states occurring concomitantly with an abrupt change in lattice parameters.[14] It should be noted when comparing the entropy of CMR materials that care must be taken to consider the moderating effect of heat capacity, which can be large.[15]

Crystallographic changes accompanying a metal-insulator transition in $R_{1-x}A_xMnO_3$ materials frequently result in interactions leading to electrical behavior that depends strongly on temperature and pressure *(P)*.[16] The exact nature of these interactions is often highly dependent on composition and structure. In addition, the electronic and magnetic properties are frequently related to the Mn-O-Mn bond angle in the perovskite structure.[17] Substitution of variously sized ions serves to distort the lattice structure and, consequently, bond angles, while also affecting the available charge carrier concentration. On the other hand, hydrostatic pressure can affect the lattice without the introduction of disorder. However, since the compressibility is anisotropic, the application of pressure results in the elongation of $MnO_6$ octahedra along the c-axis, decreasing the average Mn-O bond length and straightening the Mn-O-Mn angle.[18] A pressure study up of $\rho$ vs $T$ in polycrystalline $Pr_{0.5}Sr_{0.5}MnO_3$ showed a depression of the resistivity below $T_N$ in pressures up to 1.4 GPa.[19] Previous work has shown some variance in $\rho$ vs $T$ both in single-crystal and polycrystalline samples.[13, 14, 19] The metal-insulator transition feature in $\rho$ vs $T$ at $T_N$ is more gradual in the polycrystalline material,[13] and the onset of the FM metallic state is less pronounced,[19] or not observed.[14] In single-crystals, both the metal-insulator transition in the $\rho$ vs $T$ data and the metallic conductivity of the FM state for $T_N < T < T_C$ can be clearly observed.[6, 12, 20] Grain boundary effects in polycrystalline specimens may explain the differences seen in the $\rho$ vs $T$ data as compared to the data in single-crystals.

In order to determine the effect of hydrostatic pressure on the transport properties of $Pr_{0.5}Sr_{0.5}MnO_3$ we carried out measurements of the electrical resistivity vs temperature in a single-crystalline specimen.

**Experimental Details**

The precursor material for the crystal growth process was synthesized by solid-state reaction. Stoichiometric amounts of high purity $Pr_6O_{11}$, $SrCO_3$, and $MnO_2$ were weighed and ground together thoroughly. The mixture was placed in an alumina crucible, and reacted at 1200 °C for 22 hours in air. The sample was reground and annealed again at 1200 °C for 43 hours. The resulting powder was pressed into 5 mm diameter cylinders, which were used for the crystal growth in an optical furnace. The top and bottom stubs were rotated in opposing directions at 50 rpm, while the molten zone was swept at the rate of 7 mm/h, resulting in a 38 mm long single crystal. This growth was carried out in air at ambient pressure, after which the sample was annealed for 33 hours at 1250 °C in air, and cooled to room temperature (RT) at the rate of 0.5 °C/min. We examined samples from three cuts, located 2 mm (cut 1), 15 mm (cut 2), and 39 mm (cut 3) from the seed end. XRD analysis showed that the samples from the three cuts were single-phase, and the presence of any impurity phases could not be detected. However, the sample from cut 3 had the best Rietveld match to the I4/mcm ideal structure, yielding lattice parameters a = 5.4077 Å and c = 7.7760 Å. The larger discrepancies between the calculated and measured intensities in the samples from cuts 1 and 2 suggest that they may have more defects or non-ideal occupancies. Actually, albeit large uncertainties, a microprobe analysis showed that the sample from cut 3 had the Pr:Sr ratio closest to 1.

Magnetic susceptibility ($\chi$) and magnetization measurements in the 1.9-400 K range were conducted with the vibrating sample magnetometer option of a Quantum Design Physical Property Measurement (PPMS-9) system. Specific heat ($C_p$) measurements on a 2.7 mg piece of the crystal (cut 3) were carried out in the 1.9-300 K temperature range using a relaxation technique, in the PPMS-9 as well. Resistance measurements were carried out in the temperature range of 10-300 K with a Linear Research LR-400 four-wire bridge, using a close-cycle helium refrigerator. The measurements of $\rho$ vs $T$ in hydrostatic pressures up to 2 GPa were carried out in a self-locking piston-cylinder pressure cell. Four platinum leads were attached to the sample with silver epoxy, and the sample was mounted at the high-pressure end of a feedthrough, close to a coil of manganin which served as a manometer at room temperature. This assembly was then placed inside of a Teflon capsule filled with Fluorinert FC-75, which was used as the pressure transmitting medium, and the cell was sealed. The value of the pressure at RT was monitored with the manganin manometer. Force was applied using a hydraulic press and the pressure was locked in at RT. The small reduction in cell pressure with temperature, which is typical for this type of cell,[21] was estimated both from the changes in resistance of the manganin manometer with $T$, and from a previous characterization of the pressure loss using the superconducting transition temperature of Sn as a reference.

The temperature of the sample was monitored with a Si-diode placed on the beryllium-copper body of the cell. The cell was first cooled from RT to 10 K at a rate of ≈ 2.5 K/min. The $\rho$ vs $T$ data was collected while the cell was slowly warmed up from 10 to 300 K. In order to reduce the lag in temperature between the sensor and the sample, data between 10 K and 50 K were acquired with the refrigerator on, while a heater provided just enough power to raise the temperature of the cell at a rate ≤ 0.5 K/min. Measurements above 50 K were taken with the refrigerator and heater off, while the cell warmed naturally at a rate ≤ 0.5 K/min. The lag in temperature between the sample and the sensor was estimated to be below 0.2 K throughout the temperature range of the experiment.

**Experimental Results**

The temperature dependence of the electrical resistivity for samples from the three cuts at zero applied pressure is shown in Figure 1. As the temperature is lowered, samples of all three cuts show a drop in resistivity near $T_C$, which varied between the three samples from 255 K to 290 K, below which the resistivity continues to decrease. The drop in resistivity near $T_C$ is consistent with the suppression of spin-flip scattering due to the onset of FM order. The $\rho$ vs $T$ data for cut 3 show a clear metal-insulator transition at $T_N \approx 161$ K. In order to carry out a systematic characterization of the effect of pressure, we took both transition temperatures, $T_C$ and $T_N$, from the peaks in the derivative $d\rho/dT$. There is a significant variance in the determination of $T_N$ for these compounds in the literature, possibly reflecting how strongly dependent the crystallographic, electronic, and magnetic properties are to subtle changes in composition. For example, the low temperature transition in single-crystals is given as 141 K,[6] 150 K,[12] and 155 K,[20] while the value of $T_N \approx 161$ K in a polycrystal[7] matches the value for the single-crystal of our work. The remarkable difference in behavior between samples, even from the three cuts of the same crystal, underscores how sensitive the electronic properties are to composition. Comparison with the $\rho$ vs $T$ data in Ref. 20 suggests that the Pr:Sr ratio of the cut 1 sample may be close to 0.48:0.52. In light of the better structural and compositional properties of cut 3, we choose it for our pressure studies, as well as for the measurement of magnetization and specific heat.

Measurements of the $M$ vs $T$ for $\mu_0 H = 0.5$ T in the cut 3 specimen are shown in Fig. 2. These data show typical FM behavior slightly below room temperature, as well as the onset of the AFM phase at $\approx 161$ K. The feature in $M$ vs $T$ at $T_N$ coincides in temperature with the metal-insulator transition seen in the $\rho$ vs $T$ data of Fig. 1. The inset in Fig. 2 shows a plot of $\chi^{-1}$ vs $T$ for 200 K < $T$ < 400 K. The value of $\chi$ above $T_C$, in this case for 300 K ≤ $T$ ≤ 400 K, could be fit to a Curie-Weiss expression, $\chi = \chi_0 + C/(T-\theta)$, yielding $\theta \approx 265$ K. The value of the effective magnetic moment extracted from the $\chi^{-1}$ vs $T$ data above 350 K is $\mu_{eff} = 5.56$ $\mu_B$. Assuming that the contribution of $Pr^{3+}$ to this value is 3.58 $\mu_B$/ion, the average value of the effective moment for each Mn ion is 3.77 $\mu_B$, slightly lower than the $2[S(S+1)]^{1/2}$ value of 3.87 $\mu_B$ for $Mn^{4+}$.

The magnetization curves for this sample are shown in Fig. 3. The isotherms of $M$ vs $\mu_0 H$ clearly reflect the onset of FM order as the temperature is reduced from 300 to 250 K. Below $T_N$ the magnetization curves at $T$ = 20, 100, and 150 K clearly show that a metamagnetic transition takes place, and that the spin realignment field is inversely correlated to the temperature. The hysteretic nature of the metamagnetic transition suggests that the phase transformation from the FM to AFM phase has a first order character.

The temperature dependent specific heat $C_p(T)$ data for the cut 3 sample are shown in Fig. 4, where two remarkable features can be identified; 1) a peak is seen near 260 K, consistent with the onset of magnetic ordering at $T_C$, and 2) a double-peak feature is observed near $T_N$. As shown in the inset, the effect of a magnetic field $\mu_0 H$ = 1 T is to shift this feature at $T_N$ to lower temperatures, therefore extending the T-range of the FM order.

The temperature dependence of the electrical resistivity of the cut 3 sample was examined in hydrostatic pressures up to 2 GPa. The $\rho$ vs $T$ data for several nominal pressures at room temperature are shown in Fig. 5. The FM transition temperature near 255 K (value taken from a maximum in $d\rho/dT$ near the transition) drops gradually with pressure, reaching $\approx$ 249 K at $\approx$ 1.9 GPa. Below $T_C$ the crystal enters a FM state, where the resistivity drops with pressure as well as with cooling, as shown in Fig. 5. A transition from FM metal to AFM insulator is seen at $T_N \approx$ 161 K for the no-load measurement in the pressure cell. The application of pressure shifts this transition to higher temperatures in an almost linear fashion, increasing it from $\approx$ 161 K (no-load) to $\approx$ 188 K for $P \approx$ 1.7 GPa. The pressure dependence of the transition temperatures $T_N$ and $T_C$ is plotted in Fig. 6a, where the contours of the magnetic phase diagram are delineated. The effect of pressure on the change in resistivity at the FM-AFM phase boundary is plotted Fig. 6b. These data suggest that the metal-insulator transition could be completely suppressed at $P \approx$ 2.4 GPa.

Isotherms of $\rho$ vs $P$ are shown in Fig. 7. The ambient temperature electrical resistivity decreases approximately linearly as pressure is increased from 0 to 2 GPa. The effect of pressure on the resistivity is much stronger below $T_N$, where upon the suppression of the AFM phase in favor of the FM, the metallic state is recovered. For example, the value of $d\rho/dP$ changes from about -0.5 m$\Omega$-cm/GPa at 300 K to -1.3 m$\Omega$-cm/GPa and -1.9 m$\Omega$-cm/GPa at 100 K and 50 K, respectively, reflecting the gradual suppression of the AFM insulating state, and promotion of the conducting FM phase.

**Discussion**

In order to understand the decrease of $\rho$ and the suppression of the metal (FM) -insulator (AFM) transition at $T_N$ with pressure in $Pr_{0.5}Sr_{0.5}MnO_3$, we need to examine the effect of pressure on the lattice, and on the magnetic order. The effect of pressure on the lattice is to reduce the Mn-O bond lengths both in the FM-I4/mcm and the AFM-Fmmm phases, while straightening the Mn-O-Mn angles above $T_N$, and decreasing them below.[18] The sharp increase in $\rho$ near $T_N$ was observed to diminish with pressure both in a single-crystal (this work) as well as in a polycrystalline material.[19] However, while pressure gradually drives $\rho$ vs $T$ data below $T_N$ towards metallic behavior in the single-crystal, the polycrystalline material shows non-metallic behavior in pressures up to 1.4 GPa,[19] perhaps underlining the contribution of the grain boundaries to the transport properties.

A neutron powder diffraction study of $Pr_{0.5}Sr_{0.5}MnO_3$ under pressure reveals the gradual transformation of the AFM-Fmmm phase into the FM-I4/mcm phase below $T_N$.[18] It is argued that the metal-insulator transition results from the AFM superexchange interactions winning over the double-exchange mechanism of the FM state at $T_N$.[18] The effect of pressure below $T_N$ is then to promote the tetragonal (FM) phase at the expense of the orthorhombic (AFM) phase. At the same time, a decrease in the inter-ionic distances due to pressure possibly strengthens the superexchange interaction, therefore shifting the onset of the AFM phase to higher temperatures, and increasing $T_N$, as observed in this work. With successive gains in pressure, the FM/AFM phase ratio increases, and the discontinuity in $\rho$ at $T_N$ is gradually suppressed. Extrapolation of the $\Delta\rho$ vs $P$ data near $T_N$ suggests that the metal-insulator transition can be completely suppressed for $P \approx 2.4$ GPa. The effect of pressure on the electronic density of states (DOS) must also be considered. The DOS has been shown to vary with temperature across both $T_N$ and $T_C$, in epitaxial films of $Pr_{0.5}Sr_{0.5}MnO_3$.[22] The highest DOS of the FM phase is consistent with the higher electrical conductivity. The inducement of the FM phase below $T_N$ by pressure is therefore accompanied by the suppression of the crystallographic transformation, and an increase of the DOS at the Fermi level, consistent with the more metallic behavior of $\rho$ vs $P$ under pressure.

It has been shown that the conductivity in perovskites increases as the Mn-O-Mn bond angle is straightened.[17] From a structural point of view, it is tempting to consider that the improvement in conductivity with pressure above $T_N$ is correlated both to the shortening of the Mn-O bond lengths, and the straightening of the Mn-O-Mn angles. However, the effect of pressure below $T_N$ is more complex. The Mn-O bond lengths expand, and the Mn-O-Mn angles are reduced as the materials transitions from the FM to the AFM phase, yielding to a less conducting state. The effect of pressure is twofold. First it suppresses this transition, leading to a more conductive state. Secondly, the Mn-O bond lengths of the residual AFM regions of the mixed phase regime ($P < 2.4$ GPa) are reduced by pressure, leading to an enhancement of the superexchange interaction, and to an increase in $T_N$.

Similarly to our results, the effect of pressure on $\rho$ vs $T$ for polycrystalline $Pr_{0.5}Sr_{0.5}MnO_3$ was to raise $T_N$ towards $T_C$.[19] However, extrinsic effects due to the presence of grain boundaries broaden the transitions, which complicates the analysis of the results. The sharper features of the $\rho$ vs $T$ data on our single-crystal clearly suggest that pressure suppresses the AFM phase is favor of the FM phase.

The double-peak feature in the $C_p$ vs $T$ data correlates well in temperature with the AFM-FM phase transition. It is tempting to ascribe this double-peak feature, i.e., the ≈ 5 K separation between the two peaks, to the crystallographic and magnetic phase transformations taking place at slightly different temperatures. However, this point will require further investigation. The full shift of this feature with magnetic field suggests that it is intrinsic. The ≈ 10 K shift of the double-peak feature at $T_N$ to lower temperatures for $\mu_0 H = 1$ T is consistent with the magnetoresistance data in Ref. 6, in which the resistive anomaly at $T_N$ is shifted down in temperature by the same amount, suggesting that the effect of the magnetic field is to drive the spin realignment transition, and the crystallographic transformation to lower temperatures.

**Summary**

In summary, we have investigated the magnetic, thermal, and electrical transport properties of single-crystalline $Pr_{0.5}Sr_{0.5}MnO_3$, the latter in hydrostatic pressures up to 2 GPa. It is well established that the FM order at $T_C \approx 255$ K is mediated by a double exchange mechanism, while superexchange interactions lead to an AFM order at $T_N \approx 161$ K. The effect of pressure is to depress $T_C$ and increase $T_N$ at the respective rates of ≈ -3.2 K/GPa, and 14.2 K/GPa. Although the value of $T_N$ increases with pressure, due to the shortening of the Mn-O bonds, and resulting enhancement of the superexchange interaction, the AFM-Fmmm state is progressively suppressed, as pressure stabilizes the FM-I4/mcm phase to lower temperatures. Extrapolation of our $\Delta\rho$ vs $P$ data with pressure at $T_N$ suggests that the AFM phase should be completely suppressed near 2.4 GPa. In light of the theoretical prediction that a phase transition from A-type to C-type AFM may occur in $Pr_{0.5}Sr_{0.5}MnO_3$ near 10 GPa,[18] studies in higher pressures are in order. Similarly, the nature of the double-peak feature in the $C_p$ data near $T_N$ needs to be clarified.

*Acknowledgment*

This material is based upon work supported by the National Science Foundation Grants No. DMR-0306165 (SDSU), DMR-0244058 and DMR-0504769 (MSU).

Figure Captions

Figure 1- Normalized electrical resistivity $\rho/\rho_{300K}$ vs $T$ for samples from the 3 cuts of the $Pr_{0.5}Sr_{0.5}MnO_3$ single-crystal. The drop in $\rho$ near 255 K (cut 3) is due to the onset of ferromagnetism, while a metal-insulator transition can be observed near 161 K. For clarity, only a partial number of data points is shown. The solid lines are guides to the eye.

Figure 2- Magnetization vs temperature in $\mu_0 H = 0.5$ T for $Pr_{0.5}Sr_{0.5}MnO_3$ (cut 3). The inset shows a plot of $\chi^{-1}$ vs $T$ at high temperatures. A fit of the $\chi^{-1}$ vs $T$ data to the Curie-Weiss law for 300 K $\leq T \leq$ 400 K yields a Curie-Weiss temperature $\Theta \approx$ 265 K. The effective moment yielded from the $\chi^{-1}(T)$ data for $T \geq$ 350 K is $\mu_{eff} \approx 5.56\mu_B$.

Figure 3- Magnetization curves for $Pr_{0.5}Sr_{0.5}MnO_3$ (cut 3) for 320 K $\leq T \leq$ 20 K in magnetic fields up to 9 T. The isotherms between 250 K and 320 K are spaced by 10 K. The $M$ vs $\mu_0 H$ curves below $T_N$ show spin realignment behavior, consistent with the occurrence of a metamagnetic transition. The hysteretic behavior of the latter are reminiscent of first order phase transitions.

Figure 4- Specific heat vs temperature for $Pr_{0.5}Sr_{0.5}MnO_3$ (cut 3). The feature near 260-266 K is due to onset of FM order, while the double-peak feature centered near 145 K is due to the AFM - FM phase transformation. The latter phase change is shifted to lower temperatures in magnetic field (see inset).

Figure 5- Quasi-isobaric measurements of $\rho$ vs $T$ in pressures up to 2.0 GPa. For clarity, only a small number of data points is shown. The inset shows the detailed behavior of $\rho$ vs $T$ near the metal-insulator transition for $P \approx$ 0 (no-load), 0.8 and 1.7 GPa. The values of $T_N$ were determined from the peak in $d\rho/dT$. The $P$ values in the inset are corrected to account for the change in pressure with the cell temperature.

Figure 6- (a) Magnetic phase diagram and pressure dependence of $T_C$ and $T_N$. The values of $T_C$ and $T_N$ were taken from the maxima in $d\rho/dT$ near the transitions. The dashed line is a qualitative estimate of the phase diagram for $P >$ 2.0 GPa; (b) Effect of pressure on the change in $\Delta\rho$ at the AFM-FM transition, suggesting that the AFM phase could be suppressed near 2.4 GPa.

Figure 7- Isotherms of $\rho$ vs $P$ for $Pr_{0.5}Sr_{0.5}MnO_3$ in pressures ($P_{300K}$) up to 2 GPa. The magnitude of $d\rho/dP$ below $T_N$ is 3-4 times higher than above it, reflecting the suppression of the insulating AFM phase in favor of the metallic FM phase.

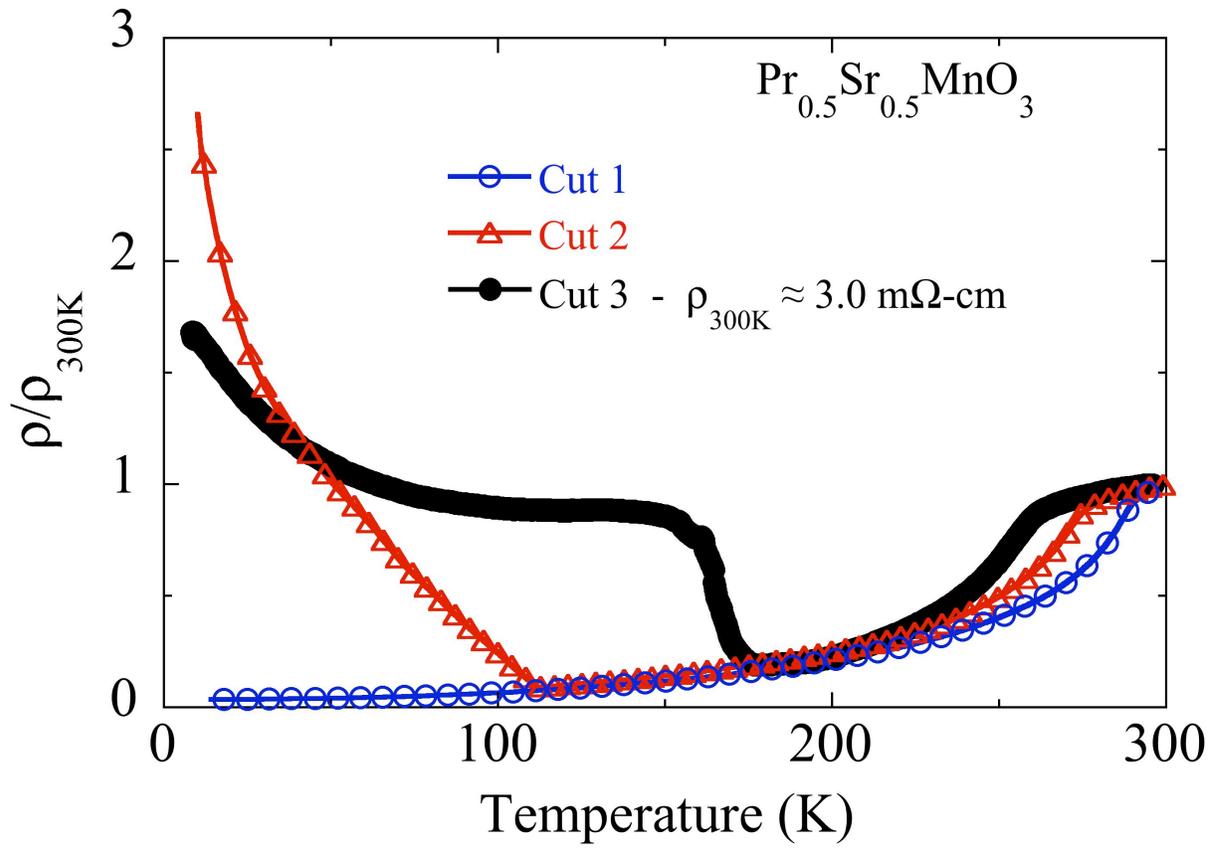

Figure 1

Rueckert et al.

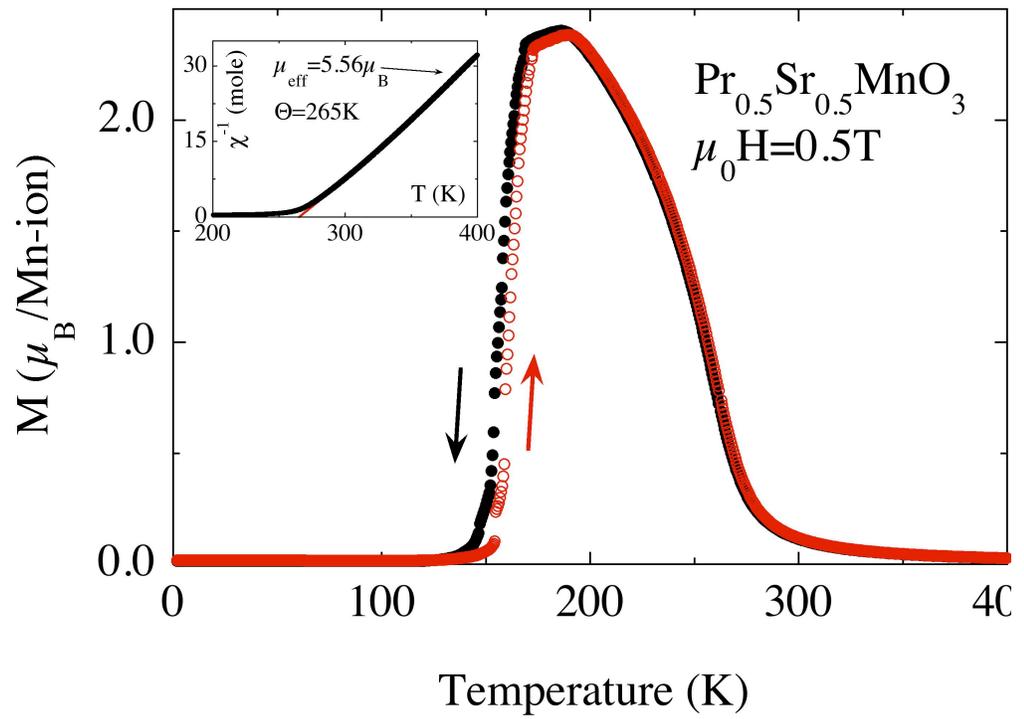

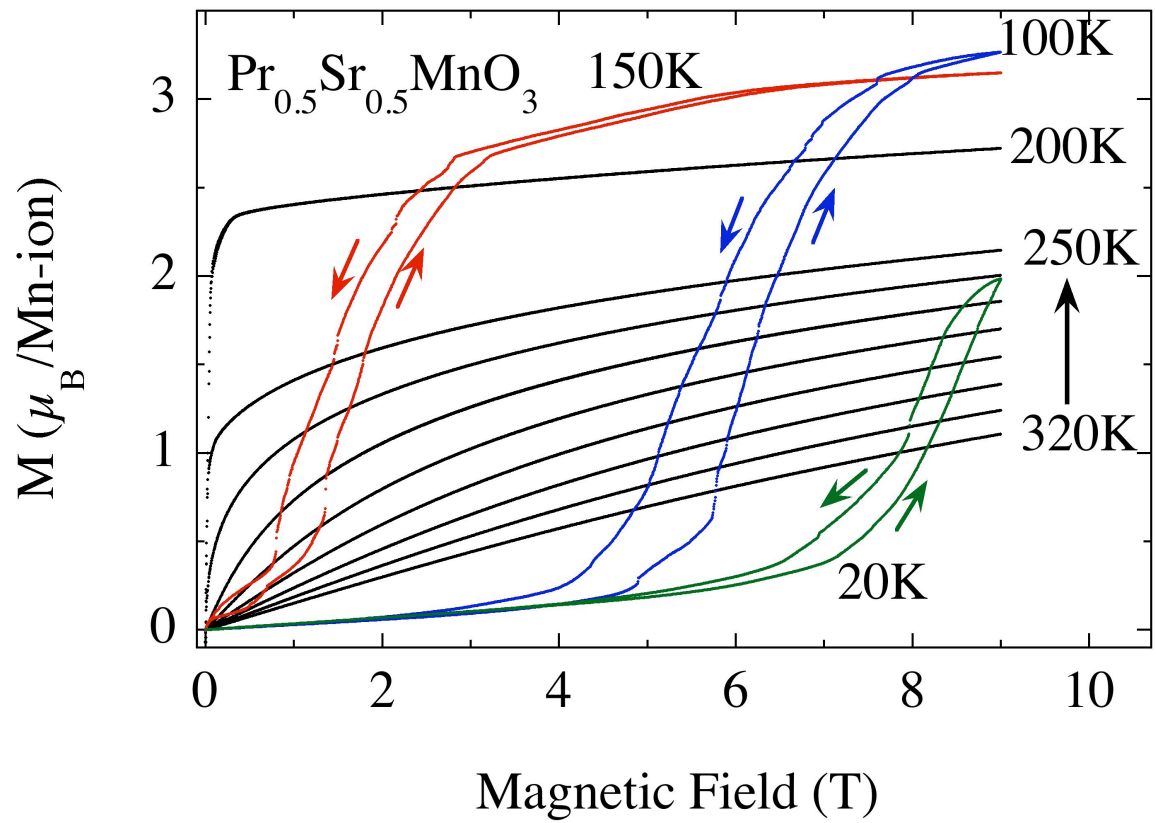

Figure 3

Rueckert et al.

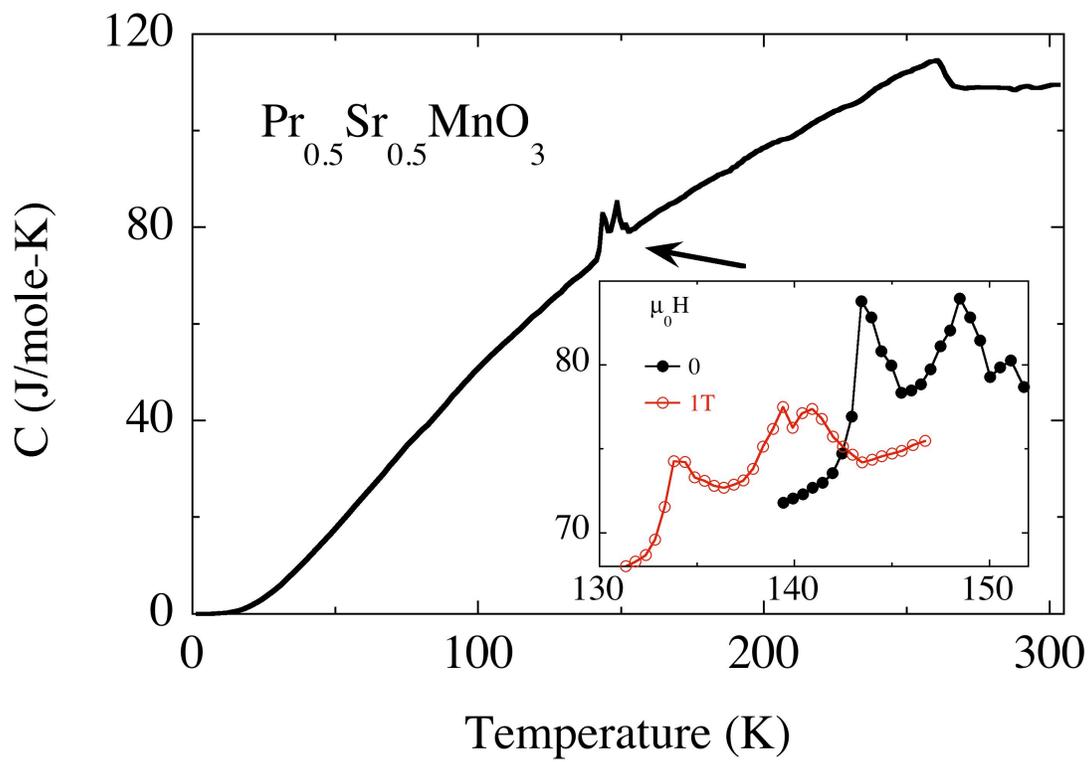

Figure 4

Rueckert et al.

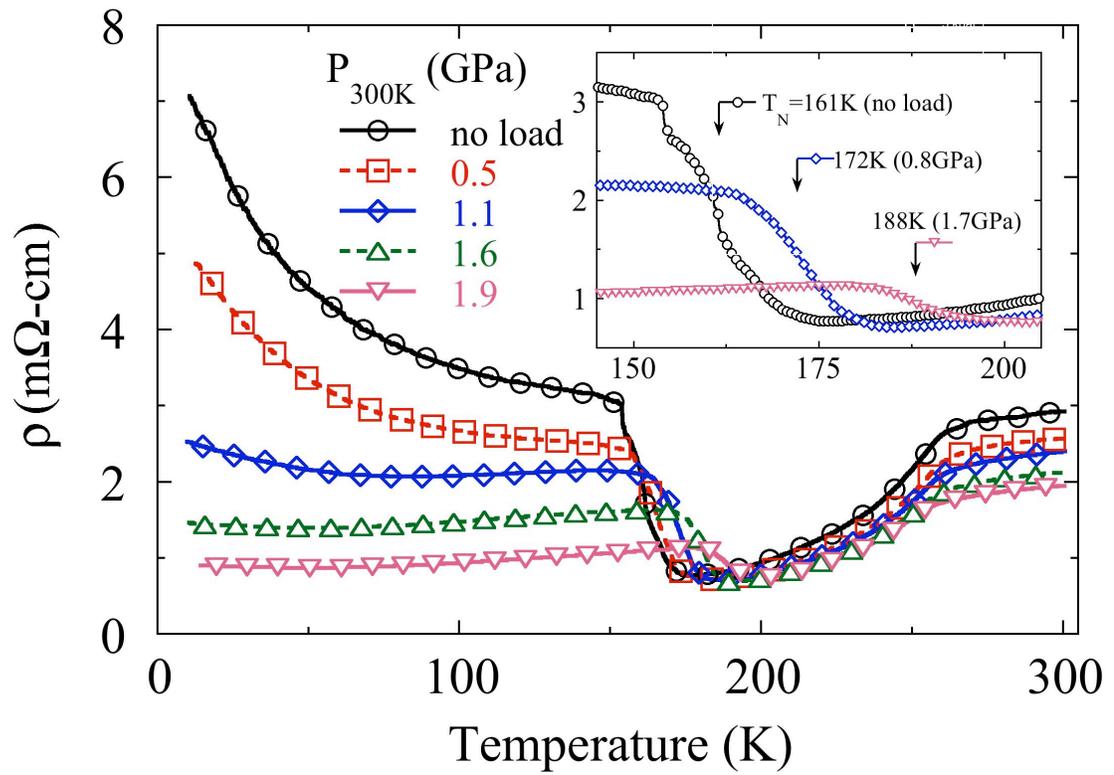

Figure 5

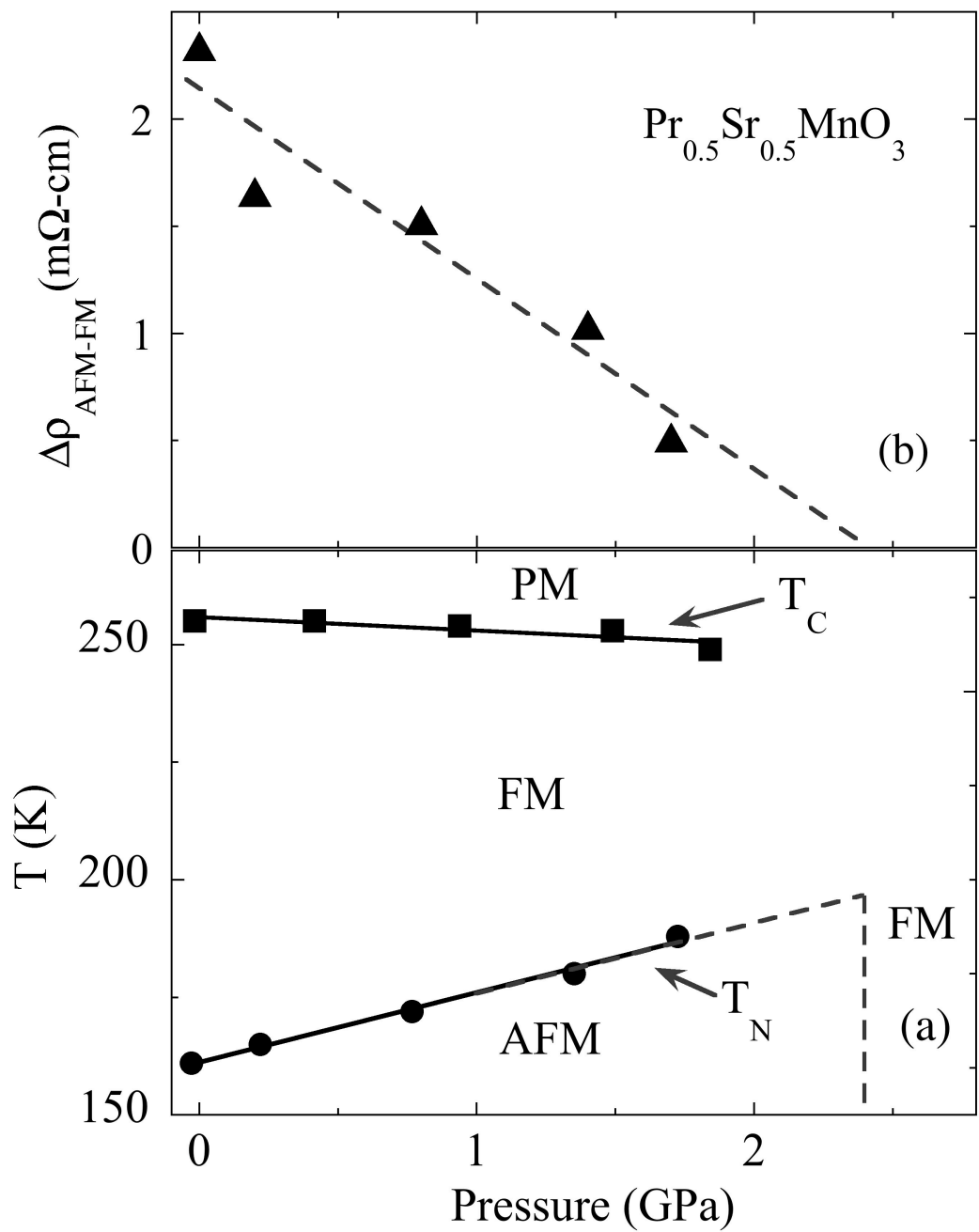

Figure 6

Rueckert et al.

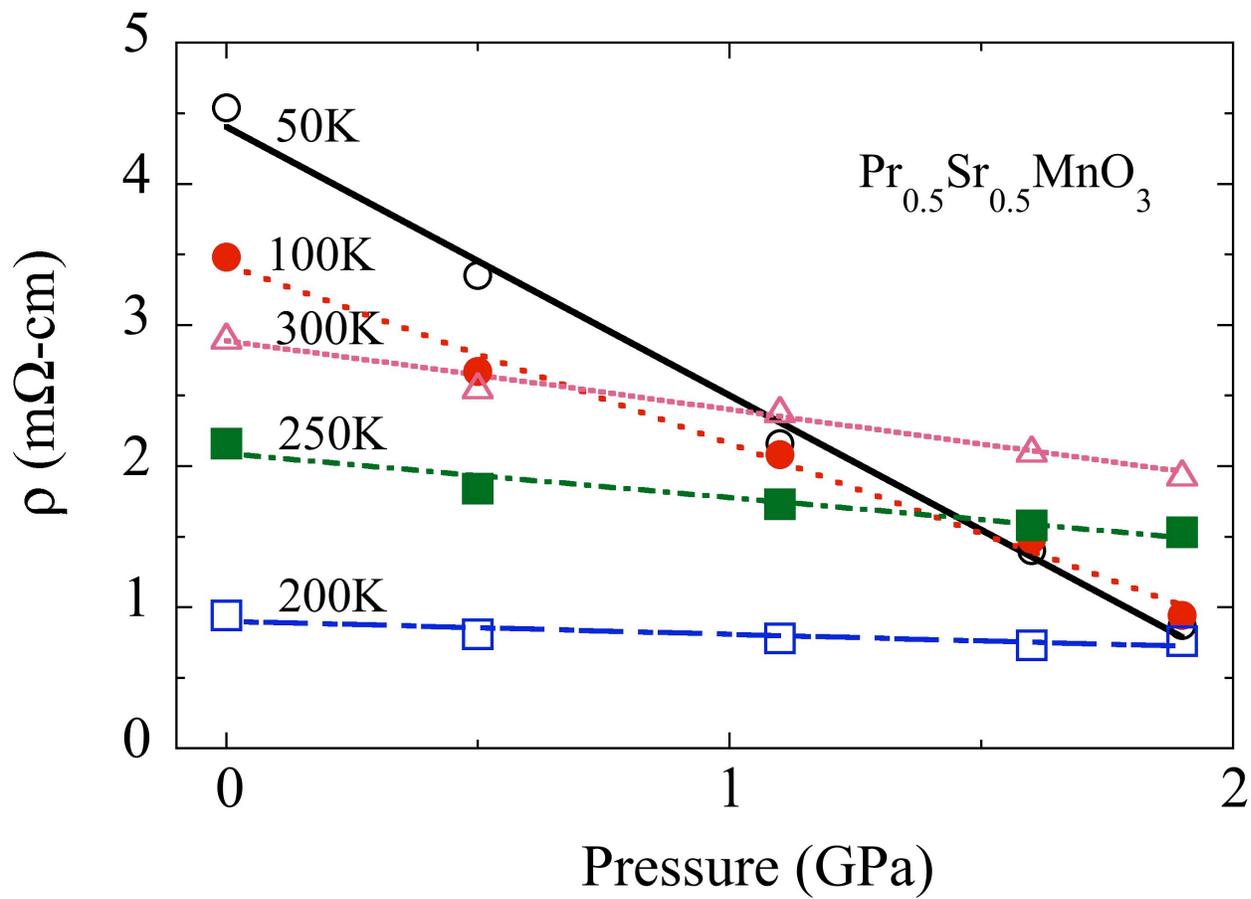

Figure 7

Rueckert et al.